\documentclass[twocolumn]{aastex631}
\newcommand{\LyA}{Ly$\alpha$\,}
\newcommand{\lya}{\LyA}
\def\la{\mathrel{\hbox{\rlap{\hbox{\lower3.6pt\hbox{$\sim$}}}\hbox{\raise1.4pt\hbox{$<$}}}}}
\def\ga{\mathrel{\hbox{\rlap{\hbox{\lower3.6pt\hbox{$\sim$}}}\hbox{\raise1.4pt\hbox{$>$}}}}}
\def\ergcm2s{\ifmmode {\rm\,erg\,cm^{-2}\,s^{-1}}\else
                ${\rm\,ergs\,cm^{-2}\,s^{-1}}$\fi}
\def\ergsec{\ifmmode {\rm\,erg\,s^{-1}}\else
                ${\rm\,ergs\,s^{-1}}$\fi}

\def\Msun{M_\odot}

\newcommand{\hii}{H{\sc i}~21\,cm}
\newcommand{\hi}{H{\sc i}}

\newcommand{\cm}{cm$^{-2}$}
\newcommand{\kms}{km~s$^{-1}$}

\shorttitle{}
\shortauthors{}

\begin{document}

\title{A Green Pea starburst arising from a galaxy-galaxy merger}
%\title{The Atomic Gas Distribution in a Green Pea Galaxy}

\author{S. Purkayastha}
\affiliation{National Centre for Radio Astrophysics, Tata Institute of Fundamental Research, Pune University, Pune 411007, India}

\author{N. Kanekar}
\affiliation{National Centre for Radio Astrophysics, Tata Institute of Fundamental Research, Pune University, Pune 411007, India}

\author{J. N. Chengalur}
\affiliation{National Centre for Radio Astrophysics, Tata Institute of Fundamental Research, Pune University, Pune 411007, India}

\author{S. Malhotra}
\affiliation{Astrophysics Division, NASA Goddard Space Flight Center, Greenbelt, MD 20771, USA}
\affiliation{School of Earth and Space Exploration, Arizona State University, Tempe, AZ 85287, USA}

\author{J. Rhoads}
\affiliation{Astrophysics Division, NASA Goddard Space Flight Center, Greenbelt, MD 20771, USA}
\affiliation{School of Earth and Space Exploration, Arizona State University, Tempe, AZ 85287, USA}

\author{T. Ghosh}
\affiliation{Green Bank Observatory, P.O. Box 2, Green Bank, WV 24944, USA}

\begin{abstract}
Green Pea galaxies are low-redshift starburst dwarf galaxies, with properties similar to those of the high-redshift galaxies that reionized the Universe. We report the first mapping of the spatial distribution of atomic hydrogen (\hi) in and around a Green Pea, GP~J0213+0056 at $z=0.0399$, using the Giant Metrewave Radio Telescope (GMRT). Like many Green Peas, GP~J0213+0056 shows strong \hii\ emission in single-dish spectroscopy, strong \lya\ emission, and a high [O{\sc iii}]$\lambda$5007\AA/[O{\sc ii}]$\lambda$3727\AA\ luminosity ratio, O32~$\approx 8.8$, consistent with a high leakage of Lyman-continuum radiation. Our GMRT \hii\ images show that the \hii\ emission in the field of GP~J0213+0056 arises from an extended broken-ring structure around the Green Pea, with the strongest  emission coming from a region between GP~J0213+0056 and a companion galaxy lying $\approx 4.7$~kpc away, and little \hii\ emission coming from the Green Pea itself. We find that the merger between GP~J0213+0056 and its companion is likely to have triggered the starburst, and led to a disturbed \hi\ spatial and velocity distribution, which in turn allowed \lya\ (and, possibly, Lyman-continuum) emission to escape the Green Pea. Our results suggest that such mergers, and the resulting holes in the \hi\ distribution, are a natural way to explain the tension between the requirements of cold gas to fuel the starburst and the observed leakage of \lya\  and Lyman-continuum emission in Green Pea galaxies and their high-redshift counterparts.

%Green Peas are starburst dwarf galaxies with properties similar to the high redshift galaxies that ionized the early universe. Despite showing strong \lya\ and LyC leakage, Green Peas are also known to be relatively gas rich. To understand how ionizing radiation escapes from Green Peas we carried out the first \hii\  mapping study of a Green Pea, GP~J0213+0056,  using the Giant Metrewave Radio Telescope (GMRT). We detect 21\,cm\, emission from an extended region around the Green Pea in a ring like structure. The strongest emission is seen from a region between GP~J0213+0056 and a companion galaxy $\sim 4.7$~kpc from the Green Pea. Our observations suggest that merger activity between GP~J0213+0056 and its companion has likely triggered the starburst and led to a disturbed \hi\  distribution and gas outflows, which in turn allowed \lya\ emission to leak from the galaxy. The scenario  seen in our results is a possible way to explain simultaneous observations of strong \lya\  emission and high \hi\  masses seen in Green Pea galaxies.

\end{abstract}

\keywords{Galaxies --- starburst, Galaxies --- dwarf, Galaxies --- 21cm line emission}

\section{Introduction} \label{sec:intro}

Faint star-forming dwarf galaxies have long been believed to be the main contributors of the Lyman-continuum (LyC) photons that ionized the Universe during the Epoch of Reionization (EoR) at $z \gtrsim 6$ \citep[e.g.][]{fan2006observational}. However, how the LyC photons escaped the EoR galaxies remains an open question today. Direct studies of these high-redshift galaxies is challenging due to limitations in both sensitivity and resolution. Studies of local analogs of the EoR galaxies, i.e. nearby galaxies with high escape fractions of LyC and \lya\ photons, hence provide an interesting avenue to understand the processes that led to cosmic reionization. 

Green Pea galaxies are extreme emission-line galaxies with compact sizes, low metallicities, and high specific star formation rates, lying at low redshifts, $z \lesssim 0.3$ \citep{cardamone2009galaxy}. They are the most extreme of the actively star-forming galaxies in the local Universe, with H$\beta$ luminosities of $\approx 10^{41} - 10^{42}$~erg~s$^{-1}$; for comparison, the ``classic'' luminous blue compact dwarf galaxy, I~Zw~18, has an H$\beta$ luminosity of $\approx 5 \times 10^{39}$~erg~s$^{-1}$, two orders of magnitude lower than that of typical Green Peas. Similarly, while galaxies of the Lyman Alpha Reference Sample \citep[LARS;][]{ostlin14} have H$\beta$ luminosities similar to those of Green Peas \citep[e.g.][]{pardy14}, eight of the 14 LARS galaxies do not meet the \lya\ rest equivalent width criterion, $\geq 20$\AA, used to identify high-$z$ \lya\ emitters (and, in fact, only one LARS galaxy, LARS14, would classify as a Green Pea). Conversely, Green Peas typically show strong \lya\ emission \citep{henry2015lyalpha,yang2016green,yang2017lyalpha}, with an equivalent width distribution matching that of high-redshift ($z\ga 3$) \lya-emitting galaxies \citep{yang2016green}. \citet{izotov2016detection,izotov2018low,izotov2018j1154+} found LyC leakage in more than ten Green Peas, with escape fractions $2-72$\%, making them the strongest known LyC leakers in the local Universe. These lines of evidence suggest that Green Peas are excellent low-$z$ analogs of the galaxies that drove cosmological reionization.

A conundrum in our understanding of Green Peas is the need for sufficient cold neutral gas to fuel the starburst and yet a low enough \hi\ column density to allow the \lya\ and LyC photons to escape. Recently, \cite{kanekar2021atomic} used the Arecibo Telescope and the Green Bank Telescope (GBT) to detect \hii\ emission from 19 Green Peas  at $z \approx 0.02 - 0.1$. Two of the Green Peas with \hii\ detections, GP~J0213+0056 at $z = 0.0399$ and GP~J1200+2719 at $z = 0.0812$, are remarkable as they show both strong \hii\ emission and strong \lya\ emission, with \lya\ escape fractions of $\approx 0.12$ and $\approx 0.39$, respectively \citep{mckinney2019neutral}. Both galaxies also have a high ratio of the [O{\sc iii}]$\lambda$5007 luminosity to the [O{\sc ii}]$\lambda$3727 luminosity, $\rm O32 \approx 8.8$ (GP~J0213+0056) and $\rm O32 \approx 12.9$ (GP~J1200+2719), suggesting a high escape fraction of LyC radiation \citep{jaskot2013origin,izotov2016detection,izotov2018low}, although we emphasize that neither galaxy has so far been detected in LyC emission. However, both galaxies are also ``gas-rich\arcsec \citep{kanekar2021atomic}, lying $> 2\sigma$ above the local relation between the \hi\ mass and the absolute B-magnitude \citep{denes2014hi}. These are currently the two best candidates to directly assess the conditions that allow the detection of both \hii\ and \lya\ emission from a single galaxy. In this {\it Letter}, we report a Giant Metrewave Radio Telescope (GMRT) \hii\ mapping study of GP~J0213+0056 at $z = 0.0399$, which has yielded the first determination of the spatial distribution of \hi\ in and around a Green Pea galaxy.

%This yielded tentative evidence for bimodality in the \hi\ properties of the sample, with $\approx 30$\% of the Green Peas lying $\pm 2\sigma$ above or below the relation between \hi\ mass and absolute B-magnitude in the local Universe \citep{denes2014new}. 

%A possible way for this to occur would be if the starburst was triggered by in-falling \hi\ through a  minor merger or a major merger with a gas-rich interacting companion galaxy. Thus, in order to understand the nature of star formation in Green Peas, it is critical to map the \hii\  emission and determine the \hi\ distribution. 

% We discuss in this paper the \hii\ mapping study of one of the galaxies with \hii\ detection, GP~J0213+0056, giving us the first direct observations of the spatial distribution of neutral atomic gas in a Green Pea galaxy. 

\section{Observations and Data Analysis} \label{sec:obs}

\begin{figure}
\includegraphics[width=3.3in]{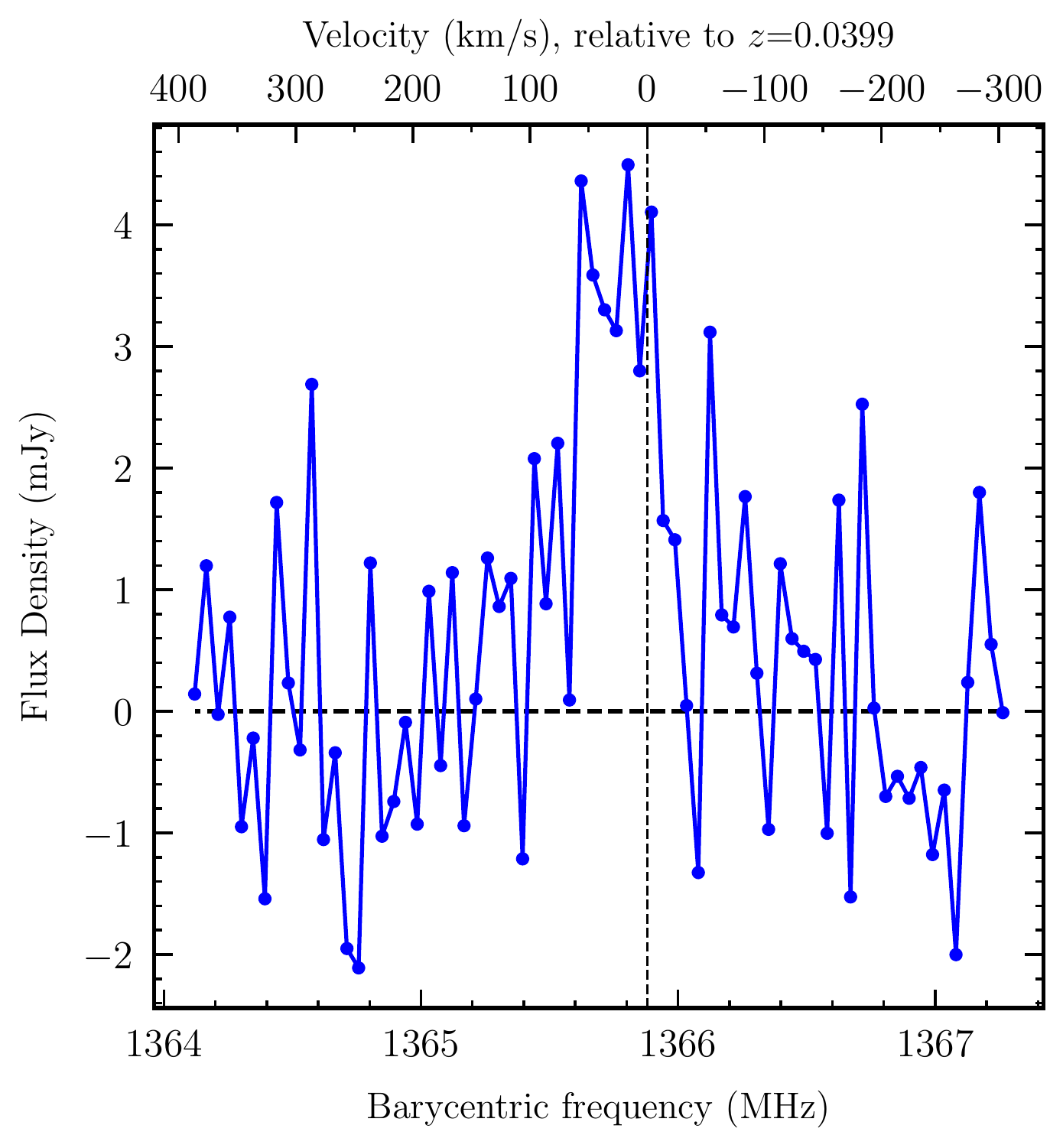}
\caption{GMRT \hii\ emission spectrum from GP~J0213+0056, from the $37\arcsec \times 32\arcsec$-resolution spectral cube. The spectrum has a velocity resolution of 10~\kms.}
\label{fig:gp0213_spec}
\end{figure}

\begin{deluxetable}{cccc}
\tablenum{1}
\tablecaption{Details of spectral cubes made at different resolutions. The columns are (1)~the FWHM of the synthesized beam used to make the spectral cube, (2)~the UV cutoff used for each cube, in kilo-wavelengths, (3)~the RMS noise at a velocity resolution of 10~\kms, and (4)~the $3\sigma$ \hi\ column density sensitivity, at a resolution of 10~\kms. \label{tab:results}}
%\tablewidth{0pt}
\tablehead{
\colhead{Beam} & \colhead{UV range} & \colhead{RMS noise} & $N_{\textsc{hi}}$\\
\colhead{($\arcsec \times \arcsec$)} & \colhead{($k \lambda$)} & \colhead{(mJy~Bm$^{-1}$)} & \colhead{($\times 10^{20}$\cm)}
}
%\decimalcolnumbers
\startdata
$37 \times 32$ & 5 & 0.98 & 0.27\\
$16 \times 16$ & 12 & 0.91 & 1.2\\
$12 \times 12$ & 13.5 & 0.87 & 2.0\\
$9 \times 9$ & 20 & 0.74 & 3.0\\
$7 \times 7$ & 25 & 0.66 & 4.4\\
\enddata
\end{deluxetable}

%3sigma Nhi sensitivity: [2.74e+19,1.18e+20,2.01e+20,3.040e+20,4.44e+20]

We observed GP~J0213+0056 (J2000 co-ordinates: 02h13m06.6s, +00d56'12.5'') with the GMRT Band-5 receivers on 20-21 December 2019 (proposal 37\_056; PI Chengalur). The observations used a bandwidth of 4.17~MHz, centred at 1373.67~MHz and sub-divided into 512 channels, with two polarizations. The GMRT Software Backend (GSB) was used as the correlator. The total on-source time was $\approx 8.4$~hours. Observations of 3C48 were used to calibrate the flux density scale and the system bandpass, while J0204+152 was used as the phase calibrator.

The data were analysed using standard procedures in the Common Astronomy Software Applications \citep[{\sc casa} Version 5.6;][]{mcmullin2007casa} package. We initially removed all visibilities from non-working antennas or affected by radio frequency interference (RFI) from the data set. We then used the calibrator data to determine the antenna-based gains and bandpass shapes, using the routines {\sc gaincalR} and {\sc bandpassR} \citep{chowdhury20}, and applied the gains and bandpass solutions to the target data. We then made a continuum image of the GP~J0213+0056 field from the line-free channels, using the routine {\sc tclean} and the w-projection algorithm. A standard iterative self-calibration procedure was then used, with a number of rounds of phase-only self-calibration and imaging, followed by inspection of the residual visibilities and further data editing to remove any data affected by RFI. The procedure was carried out until no improvement was seen in either the image or the residual visibilities on further self-calibration. The signal-to-noise ratio of the data was not sufficient to solve for both the amplitudes and the phases of the complex gains; we hence retained the gain amplitudes from the original calibration. The final continuum image has an angular resolution (full width at half maximum, FWHM, of the synthesized beam) of $2.4\arcsec \times 1.8\arcsec$ and a root-mean-square (RMS) noise of $\approx 76 \mu$Jy~Beam$^{-1}$. The positions of point sources in the GMRT image are in excellent agreement with their positions in the 1.4~GHz image of the FIRST survey \citep{becker95}, within $\approx 1.8''$ (i.e. less than half the FWHM of the FIRST synthesized beam). After applying the final gain solutions to the target data, we subtracted the cleaned continuum emission from the self-calibrated visibilities to obtain a residual visibility data set.

The unique GMRT antenna configuration, with a mix of short, intermediate, and long baselines, allowed us to make spectral cubes at a range of angular resolutions, providing information on both the total \hi\ content of GP~J0213+0056 and the spatial distribution of the \hi. For each resolution,  we first made a continuum image at the same resolution, and subtracted it out. We then made the spectral cube at this resolution from the residual visibilities. The cubes were made at resolutions of $35\arcsec$, $16\arcsec$, $12\arcsec$, $9\arcsec$, and $7\arcsec$, in the barycentric frame, with a velocity resolution of 10~km~s$^{-1}$.

%The lowest resolution spectral cube was made at the resolution of $\sim ~35\arcsec$, to match the expected \hi\ size from the \hi\ mass for GP~J0213+0056 reported by \cite{kanekar2021atomic}. We further made spectral cubes at a range of angular resolutions,  $16\arcsec,12\arcsec,9\arcsec$ and $7\arcsec$, to map the \hi\ distribution. We clean each spectral cubes with a threshold of 0.5 times the per channel rms noise to remove features in the image which might be caused by sidelobes from any remaining uncleaned components in the image. 

The cubes were made with the {\sc tclean} routine, cleaning the regions that showed \hii\ emission down to a threshold of 0.5 times the per-channel RMS noise. We used robust weighting \citep{briggs95}, with robust~$=1$ at  $35\arcsec$-resolution and robust~$=0$ for the higher-resolution cubes. This was done in order to increase the sensitivity at the lowest ($35\arcsec$) resolution in order to pick up the entire \hii\ emission, and to reduce sidelobes in the higher-resolution cubes where we aim to accurately map the \hii\ emission. Finally, we used the task {\sc imcontsub}, to fit a linear baseline to line-free channels at each spatial pixel of each cubes. We then subtracted out the fitted baselines, thus subtracting out any remaining continuum emission from each cube. 

%The \hii\ spectrum at the location of Green Pea for the lowest resolution cube is shown in Fig.~\ref{fig:gp0213_spec}~[A]. 

We used the routines {\sc immoments} in {\sc casa} and {\sc momnt} in the Astronomical Image Processing System \citep[AIPS; ][]{greisen03} to obtain velocity moments of the spectral cubes, to study the \hi\ spatial and kinematic distributions. All moment images were made from the ten velocity channels that were found by visual inspection to contain the \hii\ emission. The \hi\ spatial distribution was obtained via a simple velocity integral over the line channels of the spectral cube; this directly provides the \hi\ column density at different spatial locations. This was done for the cubes at resolutions of $7\arcsec - 16\arcsec$. For the \hi\ velocity field, we used a single  resolution, $12\arcsec \times 12\arcsec$, chosen to provide both good \hi\ column density sensitivity and good angular resolution. We used a flux density threshold of $0.7\sigma$ on the cube to exclude noise peaks, where $\sigma$ is the RMS noise on the cube per 10~km~s$^{-1}$ channel. The threshold was applied after smoothing the cube both spatially (Gaussian kernel, of FWHM $\approx 1.5$ times the synthesized beam) and spectrally (Hanning smoothing, by 10 channels). The wide Hanning-smoothing kernel was used because the spectral features are both wide and weak. We emphasize that the smoothing was done only in order to create the mask for applying the thresholds.
%We further excluded noise features in the \hi\ velocity field by blanking out the first moment image at regions that did not show a detection in the \hii\ intensity image at the same resolution.}

%To get the integrated value of the spectrum at different resolutions, we used 9 channels containing the \hii\ signal to make moment-0 maps at each angular resolution (Fig.~\ref{fig:moment0}). We further obtain the velocity field of the \hi\ around our Green Pea by making a moment 1 map (Fig.~\ref{fig:moment1}) at an intermediate angular resolution of 12\arcsec and the same channels as the moment-0 image. For both the moment-0 and moment-1 images, we use Hanning smoothing in the velocity axis to smooth over 30km/s and Gaussian smoothing in the spatial axes to smooth over the beam area. The moment images were made in Astronomical Image Processing System (AIPS).

\begin{figure*}
\centering
\includegraphics[width=6.0in]{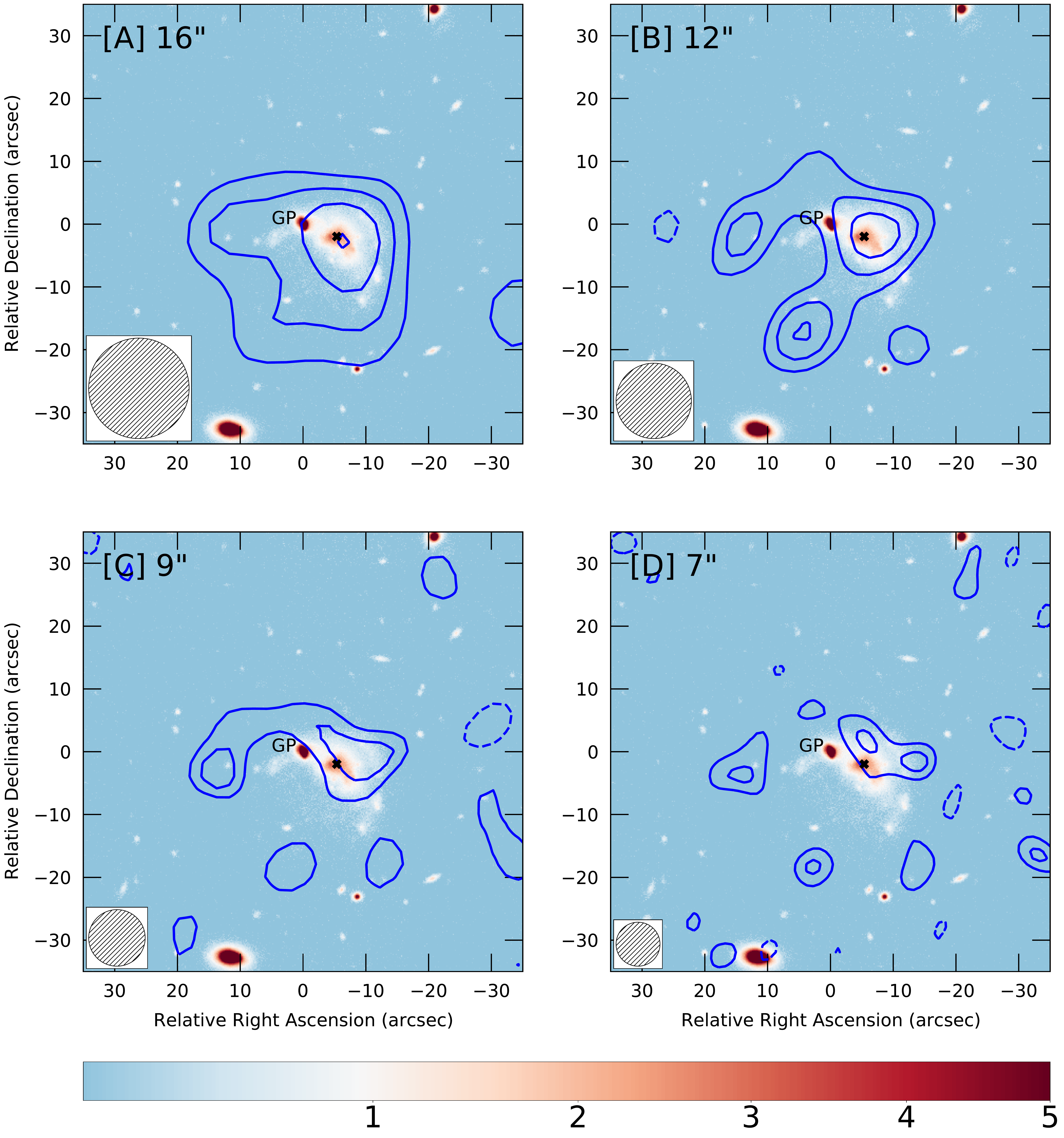}
\caption{The \hi\ spatial distribution around GP~J0213+0056 (in contours), at resolutions of [A]~$16\arcsec$, [B]~$12\arcsec$, [C]~$9\arcsec$, and [D]~$7\arcsec$, each overlaid on the Subaru HSC $i$-band image. The GMRT synthesized beam FWHM is indicated by the circle on the bottom left of each panel. The contours in each panel are at $(-2.0, 2.0, 3.0, 4.0, 5.0) \times \sigma$, with the base contours at \hi\ column densities of [A]~$1.4\times 10^{20}$~\cm, [B]~$2.2\times 10^{20}$~\cm, [C]~$ 3.5\times 10^{20}$~\cm, and [D]~$5.3\times 10^{20}$~\cm; negative values are shown with dashed contours. The locations of GP~J0213+0056 and G1 are indicated in each panel by the label ``GP'' and the symbol ``$\times$'', respectively.
} 
\label{fig:moment0}
\end{figure*}

\begin{figure}
\includegraphics[width=3.3in]{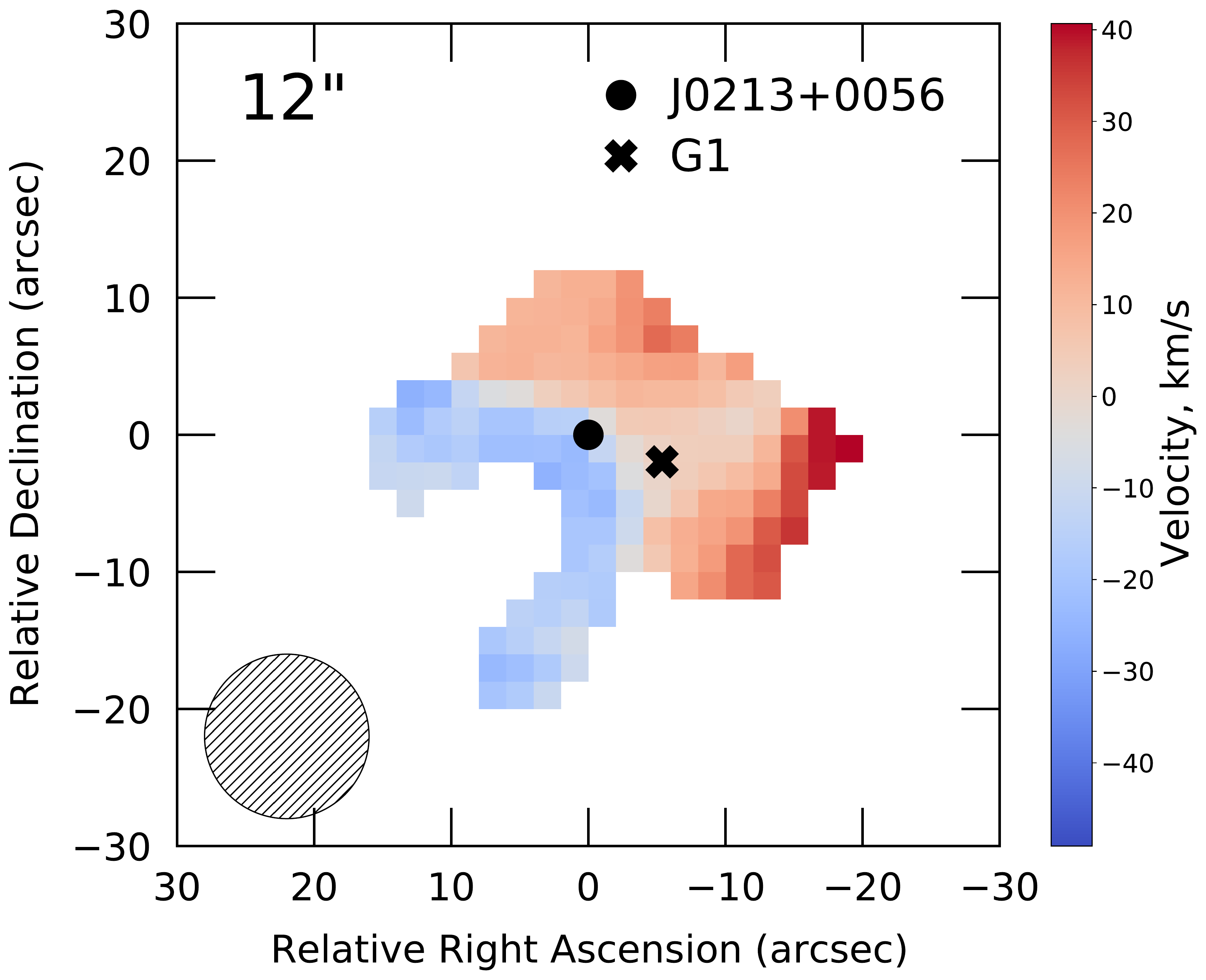}
\caption{The \hi\ velocity field of GP~J0213+0056 at an angular resolution of $12\arcsec \times 12\arcsec$; the velocity is relative to $z = 0.03995$, halfway between the redshifts of GP~J0213+0056 and G1. A velocity gradient is visible in the image, with the gas to the south and east moving towards us and the gas to the north and south-west moving away from us.}
\label{fig:moment1}
\end{figure}

%\begin{figure}
%\includegraphics[width=3.3in,trim={0cm 4cm 3.5cm 4cm},clip]{12arcsec_mom1.png}
%\caption{The \hi\ velocity field of GP~J0213+0056 at an angular resolution of $12\arcsec \times 12\arcsec$; the velocity is relative to $z = 0.03995$, halfway between the redshifts of GP~J0213+0056 and G1. The \hi\ appears to be moving away from the region between GP~J0213+0056 and G1, with the gas to the south and east moving towards us and the gas to the north and south-west moving away from us.}
%\label{fig:moment1}
%\end{figure}

\section{Results and Discussion} \label{sec:results}

Fig.~\ref{fig:gp0213_spec}[A] shows the GMRT \hii\ spectrum of GP~J0213+0056, obtained from our lowest-resolution ($37\arcsec \times 32\arcsec$) cube, at a velocity resolution of 10~km~s$^{-1}$. The RMS noise on the spectrum is $\approx 1.0$~mJy per 10~km~s$^{-1}$ channel, and the integrated \hii\ line flux density is $0.287\pm 0.044$~Jy~km~s$^{-1}$. This yields an \hi\ mass of $(2.34 \pm 0.36) \times \,10^{9} \ \Msun$, marginally lower than, but formally consistent with, the value of $(3.17 \pm 0.17)\times\,10^{9}\ \Msun$ obtained by  \cite{kanekar2021atomic} from their GBT spectrum. 

%Our GMRT images shows that the \hii\ emission in GP~J0213+0056 comes from an extended region around the Green Pea. Fig.~\ref{fig:gp0213_spec}~[A] shows the GMRT spectrum at the location of the Green Pea galaxy for the $\sim 35\arcsec$ resolution cube at a velocity resolution of 10 km/s. The spectral rms noise is $\sim 1$mJy, with an integrated \hii\ line flux density of $0.287\pm 0.044$ Jy~km/s. This gives us an \hi\ mass of $(2.34 \pm 0.36)\times\,10^{9}\Msun$, which is slightly less than the value of $(3.17 \pm 0.17)\times\,10^{9}\Msun$ reported by \cite{kanekar2021atomic}. 

%The $35\arcsec \times 35\arcsec$-resolution GMRT cube also shows evidence of \hii\ emission $\approx +4\arcmin$ east of GP~J0213+0056. The \hii\ spectrum from this companion object (hereafter, G2, at RA. 02h13m26.64s, Dec. +00d54$\arcmin$25.74$\arcsec$) is shown in Fig.~\ref{fig:gp0213_spec}[B]  and yields an \hi\ mass of $(7.1 \pm 1.3)\times\,10^{8} \ \Msun$. The GBT primary beam has an FWHM of $\approx 9'$ at the redshifted \hii\ line frequency; thus, the GBT \hii\ spectrum towards GP~J0213+0056 would contain a contribution from the emission of G2. Indeed, the sum of the \hii\  emission from GP~J0213+0056 and G2 in Fig.~\ref{fig:gp0213_spec} is in good agreement with the total \hii\ emission seen by the GBT. Sloan Digital Sky Survey \citep[SDSS; ][]{sdssdr13} images show a galaxy SDSS~J021326.64+005425.7 whose position is consistent with that of G2. However,  no spectroscopic observations are available for this object and we hence cannot conclusively associate it with G2.

Fig.~\ref{fig:moment0} shows images of the \hii\ emission in and around GP~J0213+0056 at angular resolutions of [A]~$16\arcsec$, [B]~$12\arcsec$, [C]~$9\arcsec$, and [D]~$7\arcsec$, overlaid on a Subaru HypersuprimeCam (HSC) $i$-band image \citep{Aihara22} of the field. At the coarsest angular resolution of $16\arcsec$, Fig.~\ref{fig:moment0}[A] shows that the Green Pea lies away from the peak of the \hii\ emission. At higher resolution, the \hii\ emission in panels [B] and [C] is seen to  arise from a ring-like structure around the Green Pea; the \hi\ distribution is clearly disturbed, indicating a merger system \citep[e.g.][]{hibbard01}. We hence searched the SDSS for nearby companion galaxies, and identified a galaxy SDSS~J021306.26+005609.9 (hereafter, G1) at $z = 0.0400$, just $\approx 5.7\arcsec$, i.e.  $\approx  4.7$~kpc, to the south-west of GP~J0213+0056 and at almost exactly the same redshift. The locations of the Green Pea and G1 are indicated in the four panels of Fig.~\ref{fig:moment0}. The galaxy G1 is clearly detected in the Subaru HSC image, which shows that both the Green Pea and G1 are highly distorted due to the merger.

Fig.~\ref{fig:moment0} shows that GP~J0213+0056 lies at the edge of the \hii\ emission, with no emission seen from the Green Pea location or immediately east of it in the highest-resolution images of [C] and [D]. The $3\sigma$ upper limit on the \hi\ column density at the Green Pea location is $\approx 3 \times 10^{20}$~\cm, from the 9''-resolution cube, at a velocity resolution of 10~\kms. Indeed, it is clear from Figs.~\ref{fig:moment0}[C] and [D] that the strongest \hii\ emission arises from neither the Green Pea nor G1, but from the region around them. The highest \hi\ column density in our \hii\ images, $\approx 1.9 \times 10^{21}$~\cm\ in the $7\arcsec$-resolution image of [D], arises to the west of G1 and the Green Pea. The angular offset between the highest \hi\ column density location and the centre of the galaxy G1 is $\approx 7.9''$, far larger than the positional accuracy of the GMRT images ($\lesssim 1.8''$, from the continuum image). As such, we rule out the possibility that the peak of the \hi\ column density arises from G1.

\citet{mckinney2019neutral} tentatively detect weak \lya\ absorption in GP~J0213+0056, around the much stronger \lya\ emission, obtaining an \hi\ column density of $(1.00 \pm 0.48) \times 10^{20}$~\cm. This is consistent with the observed \hi\ distribution in Fig.~\ref{fig:moment0}, which indicates that the GP lies at the edge of the \hii\ emission, with an \hi\ column density $\lesssim 10^{20}$~\cm.

%The \hi\ column densities derived by \citet{mckinney2019neutral} from \lya\ absorption for GP~J0213+0056 is $\approx (1.0 \pm 0.5) \times 10^{20}$~\cm\ , which is consistent with the \hi\ column density of the base contour in Fig.~\ref{fig:moment0}[A].

%Most of the \hi\ appears to lie around G1, with the region between the Green Pea and G1 having the highest column densities. For the 7\arcsec cube, using velocity width of 90km/s, we get the \hi\ column density of $N_{\textsc{hi}}\approx 1.6\times 10^{21}\,{\rm cm}^{-2}$ between the Green Pea and G1.

%The GMRT map also shows \hii\ emission to the north-west ($\sim\,+40\arcsec,+55\arcsec$) and south-west ($\sim\,+20\arcsec,-25\arcsec$) of the Green Pea (most clearly visible in Fig.~\ref{fig:moment0} [D]). There are no galaxies at these positions in the SDSS, so it is likely that this emission comes from \hi\ that has been driven out of GP~J0213+0056 and G1, either by the merger or in pre-merger tidal interaction. 

Fig.~\ref{fig:moment1} shows the \hi\ velocity field at an angular resolution of $12\arcsec \times 12\arcsec$. A velocity gradient is visible in the image, with the \hi\ in the arc towards the east and south-east of the Green Pea moving towards us, and that in the arc extending from north to south-west of the Green Pea moving away from us, GP~J0213+0056 and G1 are located along the boundary between the two regions. It is possible that the interaction between the galaxies has driven the \hi\ out from the Green Pea (and possibly from G1 as well); deeper \hii\ observations are needed to confirm this hypothesis.

\begin{figure}
\includegraphics[width=0.5\textwidth]{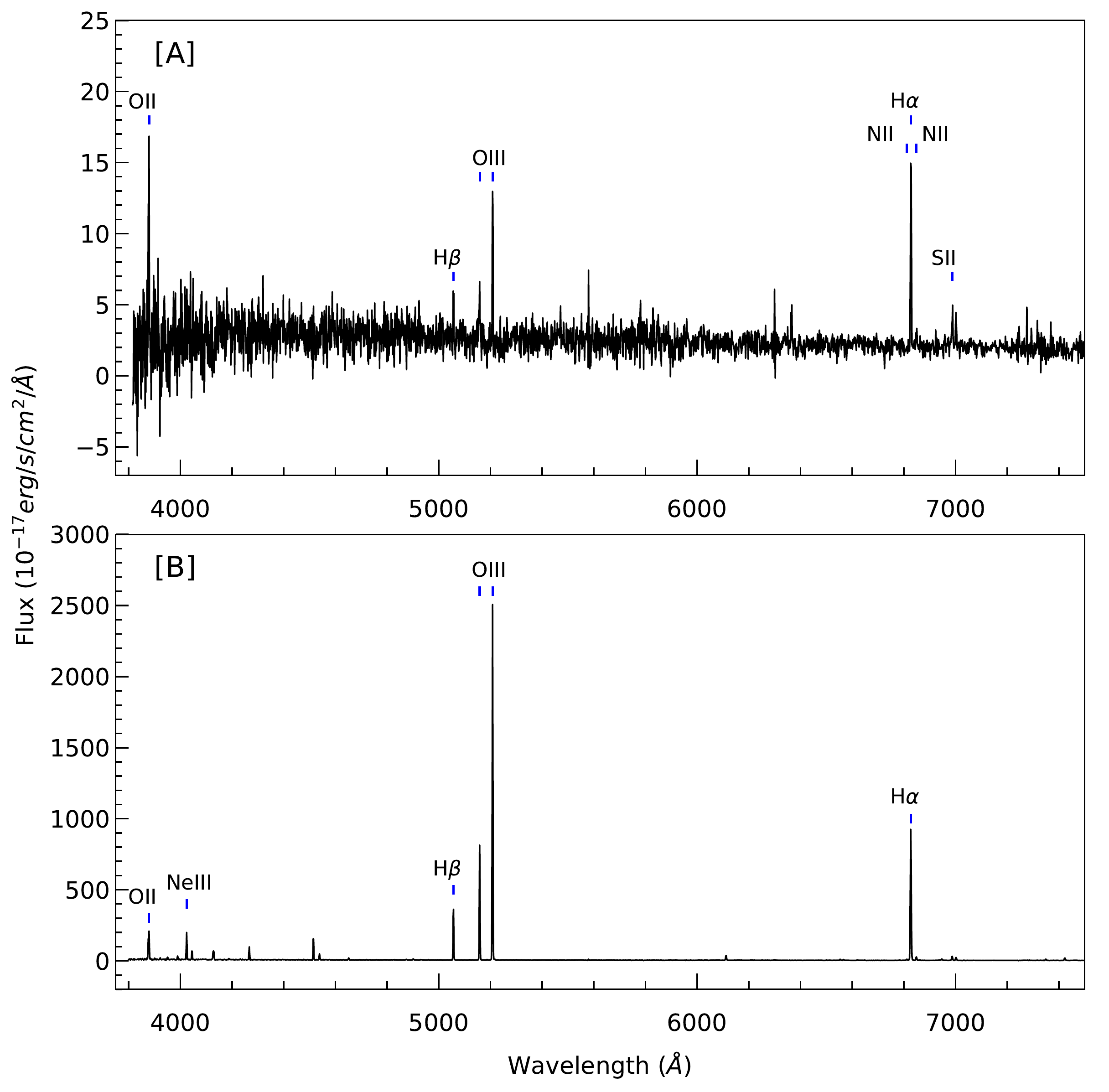}
\caption{[A] The SDSS spectrum of the galaxy G1, showing strong H$\alpha$, [O{\sc iii}]$\lambda$5007\AA, and [O{\sc ii}]$\lambda$3727\AA\ emission. [B] The SDSS spectrum of GP~J0213+0056, showing the extreme optical emission lines.} 
\label{fig:g1_spec}
\end{figure}

Fig.~\ref{fig:g1_spec} shows the SDSS spectra of G1 and GP~J0213+0056. The spectrum of G1 shows strong emission in the  H$\alpha$, [O{\sc iii}]$\lambda$5007\AA\ triplet, and [O{\sc ii}]$\lambda$3727\AA\ doublet lines, with an H$\alpha$ line luminosity of $4.8 \times 10^{40} \ \ergsec{}$, indicating active star formation. \citet{puertas2017aperture} obtain a star formation rate of $0.22 \ \Msun$~yr$^{-1}$ from the H$\alpha$ line luminosity, and a stellar mass of $\approx 1.3 \times 10^9 \ \Msun$ by modelling the optical spectral energy distribution, indicating that G1 lies on the  star-forming main sequence. We note that the stellar mass of G1 is larger than that of GP~J0213+0056 ($\approx 7.8 \times 10^8 \ \Msun$). Of course, the spectrum of GP~J0213+0056, shown for comparison, displays extreme H$\alpha$ and [O{\sc iii}]$\lambda$5007\AA\ emission, $\approx 50-100$ times stronger than that of G1.

%\section{Discussion} \label{sec:discussion}

%{\color{red}
\begin{deluxetable*}{cccccccc}
\tablenum{2}
\tablecaption{Properties of GP~J0213+0056. The columns are (1)~the ratio of the luminosity in the [O{\sc iii}]$\lambda$5007\AA\ and [O{\sc ii}]$\lambda$3727\AA\ lines, O32 \citep{kanekar2021atomic}, (2)~the equivalent width of the \LyA\ line, in \AA\ \citep{jaskot2019}, (3)~the \lya\ peak separation \citep{jaskot2017kinematics}, (4)~the \lya\ escape fraction \citep{jaskot2019}, (5)~the gas covering fraction \citep{mckinney2019neutral}, (6)~the stellar mass \citep{jiang2019correlation}, (7)~the star formation rate \citep{jiang2019correlation}, and (8)~the \hi\ mass (this work). 
\label{tab:gp_properties}}
%\tablewidth{0pt}
\tablehead{
\colhead{O32} & \colhead{EW(\lya\ )} & \colhead{$\Delta \nu_{Ly\alpha}$} &\colhead{$f^{Ly\alpha}_{esc}$} & \colhead{$f^{}_{cov}$} & \colhead{$M_{\star}$} & \colhead{SFR} & \colhead{$M_{\rm HI}$} \\
\colhead{} & \colhead{\AA} & \colhead{\kms} &\colhead{} & \colhead{} & \colhead{$10^8 \ \Msun$} & \colhead{$\Msun$~yr$^{-1}$} & \colhead{$10^9 \ \Msun$}}
%\decimalcolnumbers
\startdata
8.8 & 42 & 397 & 12\% & 0.54 & 7.8 & 1.4 & 2.34
\enddata
%\tablecomments{ }
\end{deluxetable*}
%}

 Our GMRT \hii\ mapping of the Green Pea GP~J0213+0056 has revealed that the \hii\ emission from an extended region around the galaxy, with a broken-ring structure, and evidence of disturbed kinematics, clear indications of merger activity. Consistent with the merger hypothesis, we have identified a main-sequence galaxy G1, $\approx 4.7$~kpc away from GP~J0213+0056 and at almost exactly the same redshift, $z \approx 0.0400$, with a slightly higher stellar mass. %The starburst in GP~J0213+0056 is likely to have been triggered by a major merger with the galaxy G1.
 
GP~J0213+0056 and G1 are identified as independent galaxies by the SDSS. However, the separation between the two is very small, $\approx 4.7$~kpc. It is possible that the Green Pea is a young super star-cluster formed in G1 during the major merger, i.e. that GP~J0213+0056 and G1 are not distinct galaxies. It has been suggested that such super star-clusters can form efficiently in both high-$z$ \lya-emitting galaxies   \citep{elmegreen2012globularcluster} and nearby dwarf starburst galaxies \citep{Renaud2018}. However, we note that the estimated stellar mass of GP~J0213+0056 is quite large, $\approx 7.8 \times 10^8 \ \Msun$, larger than expected for a super star-cluster within a galaxy.

GP~J0213+0056 lies at the edge of the \hii\ emission seen in Figs.~\ref{fig:moment0}[B--D], with no \hii\ emission detected at the Green Pea location in the higher-resolution \hii\ images. This lack of \hii\ emission at the location of GP~J0213+0056 suggests the presence of holes in the \hi\ spatial distribution, due to the merger. Such holes in the \hi\ distribution would facilitate the leakage of ionizing radiation from the Green Pea, consistent with the strong \lya\ emission detected by \citet{jaskot2019}, with a \lya\ escape fraction of $\approx 12 \%$. The major merger is likely to have both triggered the starburst in GP~J0213+0056 and created the pathways by which \lya, and possibly LyC, radiation could leak from the Green Pea galaxy.

Unlike other Green Peas at higher redshifts, $z \approx 0.2$ \citep[e.g.][]{izotov2018low,izotov2018j1154+}, LyC emission has not so far been directly detected from GP~J0213+0056. However, LyC escape in Green Pea galaxies has been shown to correlate with the O32 value, the gas covering fraction, and the \lya\ emission properties such as the line equivalent width (EW), the \lya\ line peak separation and \lya\ escape fraction \citep{izotov2016detection,izotov2018low,jaskot2017kinematics,jaskot2019}, although no individual property provides a definitive indicator \citep[e.g.][]{Verhamme2017}. The properties of GP~J0213+0056 are listed in Table~\ref{tab:gp_properties}. The moderately high O32 value and \lya\ equivalent width, and the moderate gas covering fraction, all suggest the possibility of LyC leakage from GP~J0213+0056. However, the \lya\ escape fraction of $\approx 12$\% is not very high and the separation between \lya\ emission peaks is large, both indicators of low LyC leakage. The current data are thus not conclusive on whether or not GP~J0213+0056 is likely to be a LyC leaker.

%This is consistent with the strong \lya\ emission detected from the Green Pea by \citet{jaskot2017kinematics}, with a \lya\ escape fraction of $\sim 11 \%$. It is also consistent with the relatively high O32 value, $\approx 8.8$, indicating LyC leakage \citep{izotov2016detection,izotov2018low}. 

%In addition, we find that the bulk of the \hi\ in the neighbourhood of GP~J0213+0056 is flowing out of the Green Pea location, also reducing the \lya\ optical depth at the systemic velocity. 

%However, galaxies showing leakage of \lya\ and LyC emission require low \hi\ column density channels through which the ionizing radiation can escape. \cite{kanekar2021atomic} gave the first estimate for the \hi\  mass in GP~J0213+0056 at a relatively high value of $(3.17 \pm 0.17)\times\,10^{9}\Msun$, which was surprising considering strong indications of \lya\ and LyC leakage. Our GMRT mapping study has shown that much of the \hi\  gas is located away from the Green Pea, with the highest column density seen in the region between GP~J0213+0056 and its companion. Further, the disturbed \hi\ distribution around the Green Pea caused by the merger might have resulted in holes in the \hi\ distribution. This would explain how \lya\ could escape the Green Pea in spite of high \hi\  mass detected from the system.

\section{Summary} \label{sec:summary}

We have used the GMRT to map the \hii\ emission from GP~J0213+0056 at $z = 0.0399$, providing the first images of the spatial distribution of \hi\ in and around a Green Pea galaxy. The GMRT images indicate that the starburst in GP~J0213+0056 is likely to have been triggered as the result of a major merger with a companion galaxy, G1, just $\approx 4.7$~kpc away, and with a stellar mass higher than that of the Green Pea. The \hii\ emission arises from a broken-ring structure around the Green Pea and G1, with no \hii\ emission detected at the location of GP~J0213+0056 in the higher-resolution images. The strong \lya\ emission observed from GP~J0213+0056 is likely to arise due to the escape of \lya\ and LyC photons through holes in the \hi\ spatial distribution caused by the merger.

\section*{Acknowledgements}
We thank an anonymous referee for a detailed report that significantly improved this manuscript.
We thank the staff of the GMRT who have made these observations possible. The GMRT is run by the National Centre for Radio Astrophysics of the Tata Institute of Fundamental Research. S.P, N.K., and J.N.C. acknowledge support from the Department of Atomic Energy, under project 12-R\&D-TFR-5.02-0700.

\bibliography{ms}{}
\bibliographystyle{aasjournal}

\end{document}